\documentclass[prl,twocolumn]{revtex4-1}

\usepackage{amsmath}
\usepackage[utf8]{inputenc}
\usepackage{braket}
\usepackage{float}
\usepackage{mathtools}

\usepackage{pgfplots}

\usepackage{hyperref}
\usepackage{lipsum}
\usepackage{xcolor}
\usepackage{comment}
\usepackage{bm,amssymb,graphicx,braket,dsfont,mathtools,bbold,MnSymbol}

\usepackage[caption=false]{subfig}
\usepackage{pgfplots}
\usepackage{cancel}
\usepackage{tabularx,booktabs}
\usepackage{multirow}
\usepackage{fancyhdr}
\usepackage{amsthm}
\usepackage{inputenc}
\usepackage{soul}
\usepackage{nccmath}

\newcommand{\proj}[1]{\ket{#1} \bra{#1}}

\newcommand{\vs}{\vec{\sigma}}

\newcommand{\abs}[1]{\left| #1 \right|}

\newcommand{\tr}{\mathrm{Tr}}
\newcommand{\lind}{\mathcal{L}}

\newcommand{\V}{\mathcal{V}}

\newcommand{\dg}{\mathcal{D}_{\gamma}}
\newcommand{\dk}{\mathcal{D}_{\kappa}}

\begin{document}
\title{Dynamics of strongly coupled disordered dissipative spin-boson systems}
\author{Eliana Fiorelli, Pietro Rotondo, Federico Carollo, Matteo Marcuzzi and Igor Lesanovsky}
\affiliation{School of Physics and Astronomy, University of Nottingham, Nottingham, NG7 2RD, UK}
\affiliation{Centre for the Mathematics and Theoretical Physics of Quantum Non-equilibrium Systems, University of Nottingham, Nottingham NG7 2RD, UK}
\date{\today}
\begin{abstract}
Spin-boson Hamiltonians are an effective description for numerous quantum many-body systems such as atoms coupled to cavity modes, quantum electrodynamics in circuits and trapped ion systems. While reaching the limit of strong coupling is possible in current experiments, the understanding of the physics in this parameter regime remains a challenge, especially when disorder and dissipation are taken into account. Here we investigate a regime where the many-body spin dynamics can be related to a Ising energy function defined in terms of the spin-boson couplings. While in the coherent weak coupling regime it is known that an effective description in terms of spin Hamiltonian is possible, we show that a similar viewpoint can be adopted in the presence of dissipation and strong couplings. The resulting many-body dynamics features approximately thermal regimes, separated by out-of-equilibrium ones in which detailed balance is broken. Moreover, we show that under appropriately chosen conditions one can even achieve cooling of the spin degrees of freedom. This points towards the possibility of using strongly coupled dissipative spin-boson systems for engineering complex energy landscapes together with an appropriate cooling dynamics.
\end{abstract}
\maketitle
\emph{Introduction---} Prominent platforms for quantum simulation, such as cavity, circuit \cite{PhysRevA.69.062320} or waveguide quantum electrodynamics \cite{PhysRevLett.111.090502} as well as trapped ions \cite{PhysRevLett.74.4091,RevModPhys.75.281} can be modeled by ensembles of two-level systems interacting via bosonic degrees of freedom (electromagnetic modes or phonons). While the weak coupling regime is relatively well understood and can be treated by a perturbative integration of the bosonic degrees of freedom, the strong coupling limit is far more challenging \cite{kockum2019ultrastrong}.

An additional layer of complexity is added by the presence of disorder, i.e. when individual spins couple to the bosonic ``environment'' at different strengths. Such a setting is relevant for at least two reasons. First, some degree of quenched disorder may always be present in realistic systems and, second, one may engineer non-uniform couplings for practical applications: systems with tunable quasi-random couplings often form the basis for a physical implementation of complex optimization problems, which may for instance be solved via quantum annealing protocols \cite{Farhi472,Denchev:PRX:2016}.

Disordered spin-boson systems have only recently moved into the focus of theoretical investigations. References \cite{Strack:PRL:2011,Goldbart:PRL:2011} explore the emergence of glassiness when many electromagnetic modes interact with an ensemble of qubits. In Refs. \cite{Rotondo:PRB:2015,Rotondo:PRL:2015}, instead, spin-glass techniques are employed to show that the same system effectively realizes an associative memory. Most of these techniques, however, cannot be straightforwardly generalized to study open quantum dynamics in the strong coupling regime, and only a few studies deal with disordered open quantum systems \cite{DallaTorre:PRA:2013,Fiorelli:PRA:2019, Rotondo:JPA:2018}. This topic acquires further relevance in the light of recent experimental progress in multimodal cavity QED, which realize tunable range \cite{PhysRevX.8.011002} and sign-changing \cite{guo2018sign} photon-mediated atomic interactions.

\begin{figure}[t]
\centering
\includegraphics[width=\linewidth]{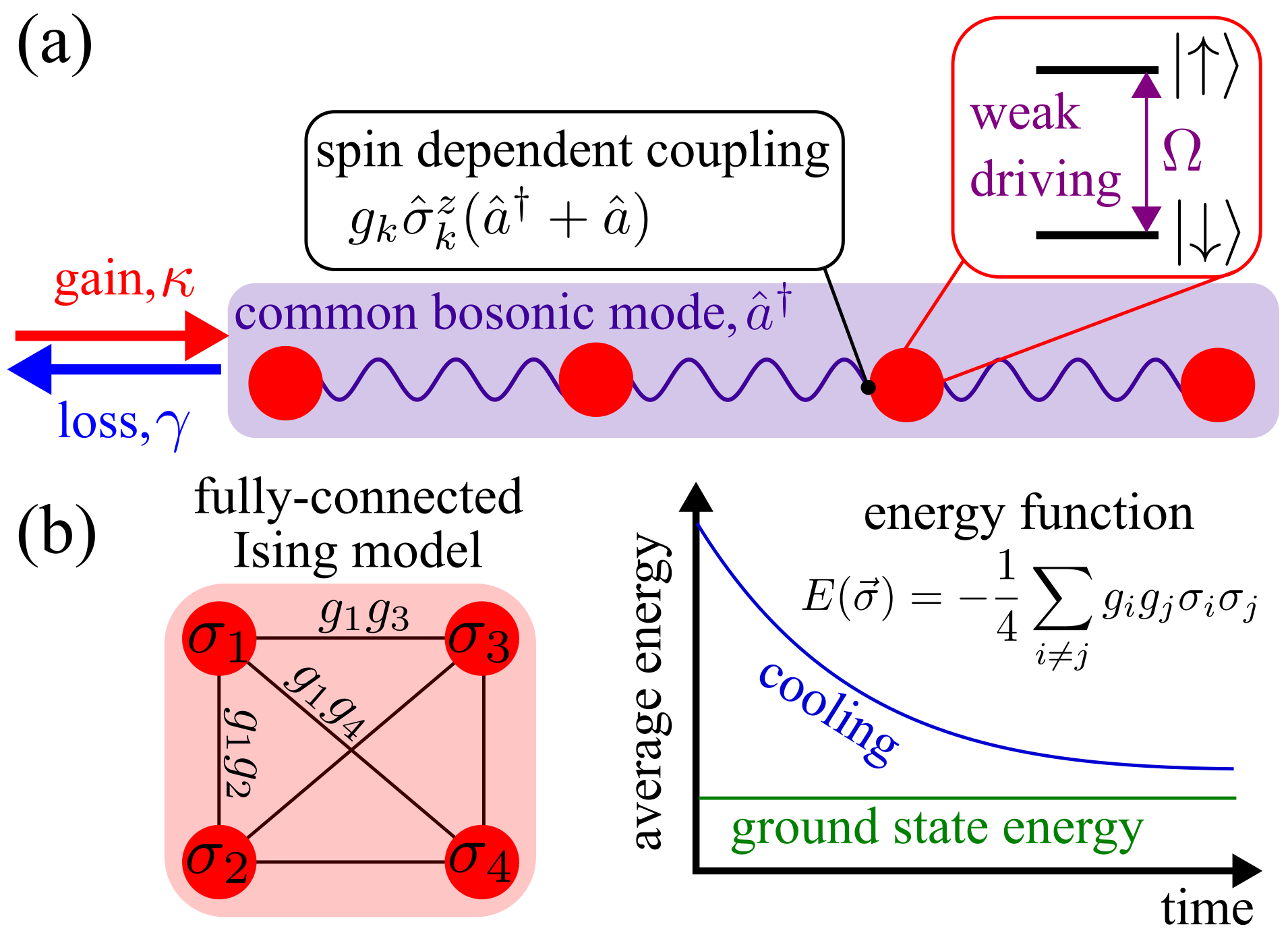}
\caption{\textbf{Dissipative spin-boson system.} (a) $N$ weakly driven (at strength $\Omega$) two-level systems (spins) are strongly coupled to a single bosonic mode with couplings $g_k$ ($k=1 , \ldots  , N$). Gain and loss of the bosons occur at rates $\kappa$ and $\gamma$, respectively. (b) The resulting effective dissipative dynamics of the spins is related to a fully-connected Ising energy function ($\sigma_k=\pm 1$), $E(\vs)$ [Eq. (\ref{cost})], in which the interaction strength between spins $i$ and $j$ is proportional to $g_i g_j$. The effective dynamics features regimes which permit cooling of the many-body spin state, i.e.~significant population of the low-energy configurations.}
\label{f:fig1}
\end{figure}

In this work we investigate a disordered and dissipative system in which weakly driven spins are strongly coupled to a bosonic mode (see Fig.~\ref{f:fig1}a). We employ a perturbative approach which relies on the weakness of the driving rather than of the spin-boson coupling. We find that the effective spin dynamics is governed by a rate equation that depends on a fully-connected Ising energy function as sketched in Fig.~\ref{f:fig1}b. Depending on the rates of bosonic loss and gain we identify several distinct dynamical regimes: two of them are high-temperature ones, in which the stationary state of the system is fully mixed. A further one mimics an effective low-temperature dynamics, which permits cooling of the spin system. Outside these the dynamics is generally non-thermal and detailed balance is broken.
This link between an open, strongly coupled spin-boson system and the physics of disordered Ising spin systems opens up the possibility of engineering complex classical energy landscapes --- with importance in the context of optimization problems \cite{Barahona_1982} or associative memories \cite{Hopfield:1982} --- together with a cooling protocol.

\emph{Model ---} We consider an ensemble of $N$ two-level systems interacting with a single bosonic mode described by the following Dicke Hamiltonian \cite{Garraway11,Kirton:AQT:2019,HeppL73,HeppL:PRA:73}:
\begin{equation}\label{e.Dicke}
\hat{H}=\omega\hat{a}^{\dagger}\hat{a}+\sum_{i=1}^N g_{i}\hat{\sigma}_{i}^{z}(\hat{a}^{\dagger}+\hat{a})+ \Omega\sum_{i=1}^N \hat{\sigma}_{i}^{x}.
\end{equation}
Here, $\hat{\sigma}_{i}^{x,y,z}$ are the Pauli operators and $\hat{a}$ and $\hat{a}^{\dagger}$ the bosonic annihilation and creation operators. The parameters $\omega$ and $\Omega$ denote the fundamental frequency of the bosons and the coherent coupling strength between the two spin states, respectively. The spin-boson couplings $g_i$ are assumed to be independent and randomly distributed with zero mean and variance $g^2$.

We include dissipation on the boson in the form of Markovian gain and loss processes. The density matrix $\rho$ of this open quantum system therefore evolves under a Lindblad equation
\begin{equation}\label{e.Lindblad}
\dot{\rho}=\lind \rho=-i [\hat{H}, \rho] +\sum_{n=l,g} \hat{L}_{n}^{\phantom{\dagger}}\rho \hat{L}_{n}^{\dagger}-\frac{1}{2} \lbrace \hat{L}_{n}^{\dagger} \hat{L}_{n}^{\phantom{\dagger}}, \rho \rbrace
\end{equation}
with the jump operators $\hat{L}_{l}=\sqrt{\gamma}\hat{a}$, $\hat{L}_{g}=\sqrt{\kappa}\hat{a}^{\dagger}$ where $\gamma$ ($\kappa$) is the loss (gain) rate and $\gamma > \kappa \geq 0$.

A physical realization of this model can for instance be achieved on trapped-ion quantum simulators \cite{PorrasC:PRL:04, AedoL:PRA:18}:
Following the scheme represented in Fig.~\ref{f:fig1}, such system would consist of $N$ ions coupling to the centre-of-mass phonon mode. As it has been shown for the quantum Rabi model \cite{mezzacapo2014digital} and eventually generalized to the Dicke model \cite{AedoL:PRA:18}, the application of multiple laser fields on the ions yields both the spin dependent coupling $g_{k}\hat{\sigma}_{k}^{z}(\hat{a}+\hat{a}^{\dagger})$ and the weak driving term, $\Omega\hat{\sigma}_{k}^{x}$ entering Eq.~\eqref{e.Dicke}. Finally, as illustrated in Fig.~\ref{f:fig1}a, the gain and loss dynamics can be achieved by applying lasers on the ions on the edge of the chain, which is discussed in Ref. \cite{lin2009large}. Since this ion is coupled to the same phonon mode as the other ions this effectively implements jump operators of the form introduced in Eq.~(\ref{e.Lindblad}).

\emph{Spin dynamics at strong coupling ---} We explore the dynamics \eqref{e.Lindblad} in the strong coupling regime, i.e.~when the driving acting on the spins is much weaker than the spin-boson interaction ($\Omega \ll g$). In the following, we sketch the perturbative technique we employ for this purpose. First, we split the Lindblad superoperator according to $\lind = \lind_0 + \lind_1$, where $\lind_1(\cdot)=-i\Omega[\sum_{i}\hat{\sigma}_{i}^{x},\cdot]$ can be regarded as a small perturbation. Focusing now on $\lind_0$, we notice that each $\hat{\sigma}_i^z$ commutes with all jump operators and Hamiltonian terms in it, implying that the $z$-components of the spins constitute $N$ independent conserved quantities \cite{Albert2014}. Hence, the dynamics can be separated in $2^N$ independent sectors labeled by the classical spin configurations $\vec{\sigma} = (\sigma_1, \ldots , \sigma_N)$ ($\sigma_i \in \set{-1,1}$), where $\hat{\sigma}_i^z \ket{\vec{\sigma}} = \sigma_i \ket{\vec{\sigma}}$; in other words, states belonging to different sectors never mix under the action of $\lind_0$. In each sector, the bosonic mode evolves according to a Lindbladian $\lind_0 (\hat{\sigma}^z_i \to \sigma_i)$ which describes a damped quantum harmonic oscillator with a (spin-configuration-dependent) spatial displacement. This admits a single (bosonic) stationary state, denoted by $\rho_{\vec{\sigma}}$. 
%
%
%
%
%
We assume that, due to the random and independent nature of the couplings $g_i$, no additional symmetries are present which could protect more complex subspaces. Hence, for any initial state $\rho_0$ of the spin-boson system the corresponding stationary state under $\lind_0$ is of the form $\rho_{\mathrm{stat}}=\sum_{\vec{\sigma}}p_{\vec{\sigma}}\rho_{\vec{\sigma}} \otimes\ket{\vec{\sigma}}\bra{\vec{\sigma}}$, where the coefficients $p_{\vec{\sigma}}$ form a set of classical probabilities.

The perturbation $\lind_1$ couples sectors corresponding to different classical spin configurations $\vec{\sigma}$. Its action can be incorporated perturbatively \cite{Degenfeld14, Marcuzzi:JPA:2014} as long as $\Omega$ is small compared to the typical rate at which coherences between sectors decay (estimated further below). We proceed by projecting onto the stationary manifold of $\lind_0$ via $P \rho(t) = \sum_{\vec{\sigma}} \tr_B \left\{ \bra{\vec{\sigma}} \rho(t) \ket{\vec{\sigma}} \right\} \rho_{\vec{\sigma}} \otimes \proj{\vec{\sigma}}$, where $\tr_{B}$ denotes the partial trace over the bosonic mode. This reduces the dynamics to the evolution of the classical probabilities $p_{\vec{\sigma}}(t) = \tr_B \left\{ \bra{\vec{\sigma}} \rho(t) \ket{\vec{\sigma}} \right\}$ according to a master equation $\dot{p}_{\vs}=\sum_{\vs'}W_{\vs' \rightarrow \vs}p_{\vs'}-W_{\vs \rightarrow \vs'}p_{\vs}$. Here $W_{\vs' \rightarrow \vs}$ is the rate for switching from configuration $\vs'$ to $\vs$. Note, that up to second order in $\Omega$, the corresponding stochastic process includes only single spin flips (i.e., $W_{\vs' \rightarrow \vs} \neq 0$ only if $\vs$ and $\vs'$ differ by a single spin). The rates read
\begin{figure*}
\centering
\includegraphics[width=\linewidth]{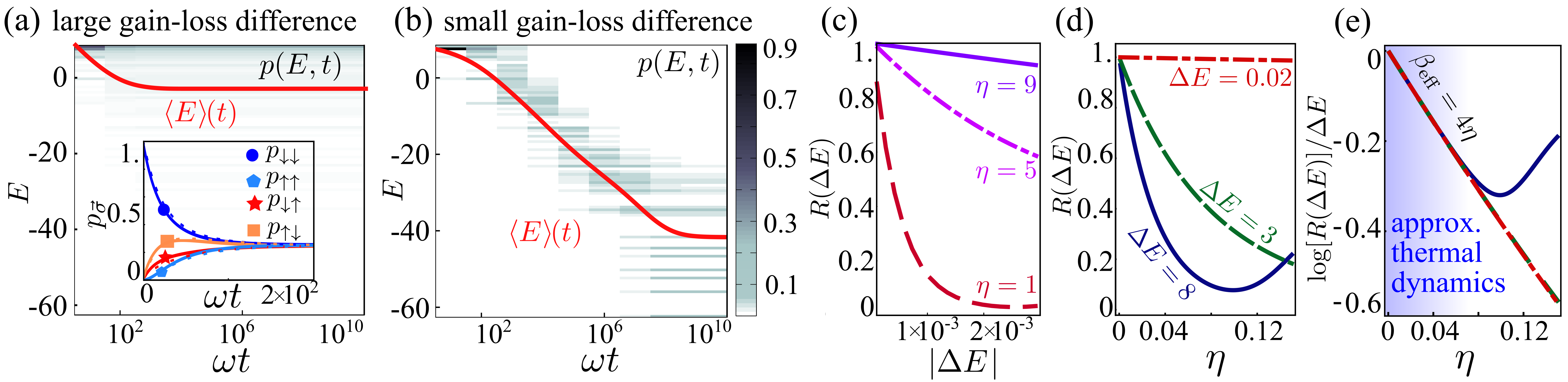}
\caption{\textbf{Regimes of effectively thermal spin dynamics.} (a) Statistical energy distribution versus time for $N = 10$ spins for large gain-loss difference $\eta$, starting from an initial state where all spins point down. The superposed red solid line displays the evolution of the average energy, $\braket{E}(t)$ as a function of time, whereas the shading represents the probability of being in a configuration with energy $E$ at time $t$. We set $\Omega=0.1$, $\kappa=1$, $\omega=1$, $\eta=10$ ($\theta = 1/11$) and we select the couplings $g_i$ from a uniform distribution in $[-g_0,g_0]$ with $g_0 = 3$. Inset: evolution of the probabilities $p_{\sigma}$ for a system of two spins with the same parameters. We compare the effective model (solid lines) with the numerically-exact diagonalization of the full open quantum problem (dashed lines), highlighting good agreement. (b) Statistical energy distribution versus time for $N = 10$ spins and corresponding evolution of the average energy, $\braket{E}(t)$ (red solid line) for $\eta = 0.1$ ($\theta = 1/20$). The dynamics clearly tends to preferentially populate the low-energy configurations at long times. (c) Ratio $R_i(\Delta E) = W_i(\Delta E) / W_i(-\Delta E)$ versus energy difference $|\Delta E|$. For small values of $\eta$ the rate $W_i(\Delta E)$ is strongly asymmetric with respect to $\Delta E$, whereas for large $\eta$ we have $W_i(\Delta E) \approx W_i(-\Delta E)$. (d) The ratio $R_i(\Delta E)$ is shown for three different values of $\Delta E = 0.02,3,8$ and the values of $g_k$ are drawn from a uniform distribution $[-g_0,g_0]$ with $g_0 = 6$. At small $\eta$, the ratio $R_i(\Delta E)$ approaches one, indicating that configurations are visited with equal probability as for large $\eta$. (e) The three curves in (d) are rescaled according to $\log R_i(\Delta E)/\Delta E$. Their asymptotic collapse in the limit $\eta \to 0$ highlights the existence of a unique inverse temperature $\beta_{\rm eff}$ which governs the dynamics when $\eta$ is sufficiently small. }
\label{f:fig2}
\end{figure*}
\begin{align}
&W_{\vs \rightarrow \vs'} = \frac{2\Omega^{2}}{\omega}\int_{0}^{\infty}\! d\tau e^{-\frac{2g_{i}^{2}\nu}{\omega^{2}}\left( f(\tau)+\tau \right) } \begin{medsize}  \cos{\left[ 16 \frac{\Delta E_{i} \tau - g_{i}^{2}s(\tau) }{\omega^{2}(\eta^{2}+4)} \right]} \end{medsize}\,,\nonumber \\
&f(\tau) = \frac{8-2\eta^2}{\eta \left(\eta^2+4\right)} \left[ 1-e^{-\frac{\eta}{2}\tau}\cos( \tau) \right]-\frac{8 e^{-\frac{\eta}{2}\tau} }{\eta^{2}+4}\sin( \tau)\,, \nonumber\\
&s(\tau) = \frac{  4\eta\left[ e^{-\frac{\eta}{2}\tau}\cos(\tau)-1 \right] + \left[\eta^{2}-4 \right]e^{-\frac{\eta}{2}\tau}\sin(\tau)  }{ \eta^{2}+4},
\label{eqd}
\end{align}
where the index $i$ denotes which spin is being flipped and changes sign between configurations $\vec{\sigma}$ and $\vec{\sigma}'$.

In Eq.~(\ref{eqd}) we have introduced the (scaled) difference between loss and gain rates $\eta=(\gamma-\kappa)/\omega \equiv \gamma/\omega(1-\theta)$, the ratio $\theta = \kappa/\gamma \in [0,1)$ and the parameter $\nu=4(1+\theta)\eta/[(\eta^2+4)(1-\theta)]$. Importantly, the sole dependence on the spin configuration is through the quantity $\Delta E_i~= g_i \sigma_i\sum_{l\neq i}g_{l}\sigma_{l}$, which can be interpreted as an energy difference (see further below). Note, that there is a characteristic scale of exponential suppression of the integrand of $W_{\vs \rightarrow \vs'}$. This corresponds to the typical timescale involved in the loss of coherence between sectors belonging to different classical spin configurations. Since the function $f(\tau)$ is bounded, we can estimate this timescale as $t_\mathrm{L} \approx \omega / (2g^2 \nu)$. Our perturbative expansion thus holds as long as $\Omega \ll 1/t_\mathrm{L}$. In the following we perform a detailed investigation of the effective spin dynamics. It turns out that the loss-gain parameter $\eta$ is central in determining the qualitative dynamical behavior: we will identify an effective high-temperature regime in the asymptotic limits $\eta \to \infty$ and $\eta \to 0^+$. Furthermore, we find an effective low-temperature (cooling) dynamics when $\eta < 1$.

\emph{Large $\eta$: infinite temperature dynamics ---} As remarked above, the quantity $\Delta E_i$ can be interpreted as the change in the energy function
\begin{equation}
E(\vs) = -\frac{1}{4} \sum_{i \neq j} g_i g_j \sigma_i \sigma_j\,.
\label{cost}
\end{equation}
occurring when the $i$-th spin is flipped, i.e.~$\Delta E_i = E(-\sigma_i) - E(\sigma_i)$. In passing, we remark that the energy levels defined by Eq.~\eqref{cost} are (at least) doubly degenerate, since $E(\vec{\sigma}) = E(-\vec{\sigma})$.
%
%
For a large gain-loss difference, $\eta \gg 1$, we find that in Eq.~(\ref{eqd}) $f(\tau) \sim s(\tau) \sim O(1/\eta)$. Therefore both functions are approximately zero and the parameter $\nu \approx 4 \tfrac{1+\theta}{1-\theta} \eta^{-1}$ determines the leading behavior of the timescale $t_\mathrm{L}$. The validity of the perturbative requirement thus imposes an upper bound to loss-gain difference, which must satisfy $1 \ll \eta \ll \tfrac{4g^2 (1+\theta) }{\omega \Omega (1-\theta)}$.

With the above approximations the rate $W_{\vs \rightarrow \vs'}$ acquires a considerably simpler form: having neglected $s(\tau)$, it no longer depends on the sign of $\Delta E_i$, implying that the rates for inverse processes $\sigma \to \sigma'$ and $\sigma' \to \sigma$ are equal. This gives rise to an infinite-temperature dynamics which populates all configurations uniformly. This behavior is highlighted in Fig.~\ref{f:fig2}(a): we show that the average energy $\braket{E}(t)$ approaches (up to finite size corrections) zero, indicating a equal population of all spins states at stationarity.

Interestingly, for large $\eta$ and up to second order in perturbation theory, the rate $W_{\vs \rightarrow \vs'}$ is formally equivalent to the dissipative dynamics of a fictitious transverse field Ising model. The corresponding Hamiltonian is $\hat{H}_{\rm eff} = \Omega_{\rm eff} \sum_i \hat{\sigma}^x_i + \xi E(\hat{\sigma}^z)$ [with $\Omega_{\rm eff} = \Omega \lambda $, $\xi = 8\lambda^2 / (\omega \eta^2)$] and the spins are subject to strong dephasing at a (site-dependent) rate $\gamma_{\rm eff , i} = \tfrac{8 g_i^2 \lambda^2 (1+\theta)}{\omega \eta (1-\theta)}$ \cite{Everest2016}. Here $\lambda$ is an arbitrary factor that should be chosen consistently with the (perturbative) requirement $\Omega_{\rm eff} / \gamma_{\rm eff,i} \ll 1$. Therefore, in this limit, the bosons can be interpreted as forming an infinite temperature bath causing dephasing of the spin degrees of freedom.

\emph{Small $\eta$: approximate low-temperature dynamics ---} For $\eta < 1$ there exists a regime in which the rate equation dynamics mimics a thermal process with finite temperature. To be precise, this limit is achieved by fixing the parameters $\kappa$ (gain) and $\omega$, while $\gamma$ (loss) is varied. Accordingly, the limit $\eta \to 0^+$ has to be interpreted as $\gamma \to \kappa^+$, so that $\theta = (1 + \eta \omega /\kappa)^{-1} \to 1$ remains finite. Provided the parameters are chosen carefully, this leads to the cooling of the spins with respect to the Ising energy function \eqref{cost} [see Fig.~\ref{f:fig2}b].

\begin{figure}
\centering
\includegraphics[width=0.6\linewidth]{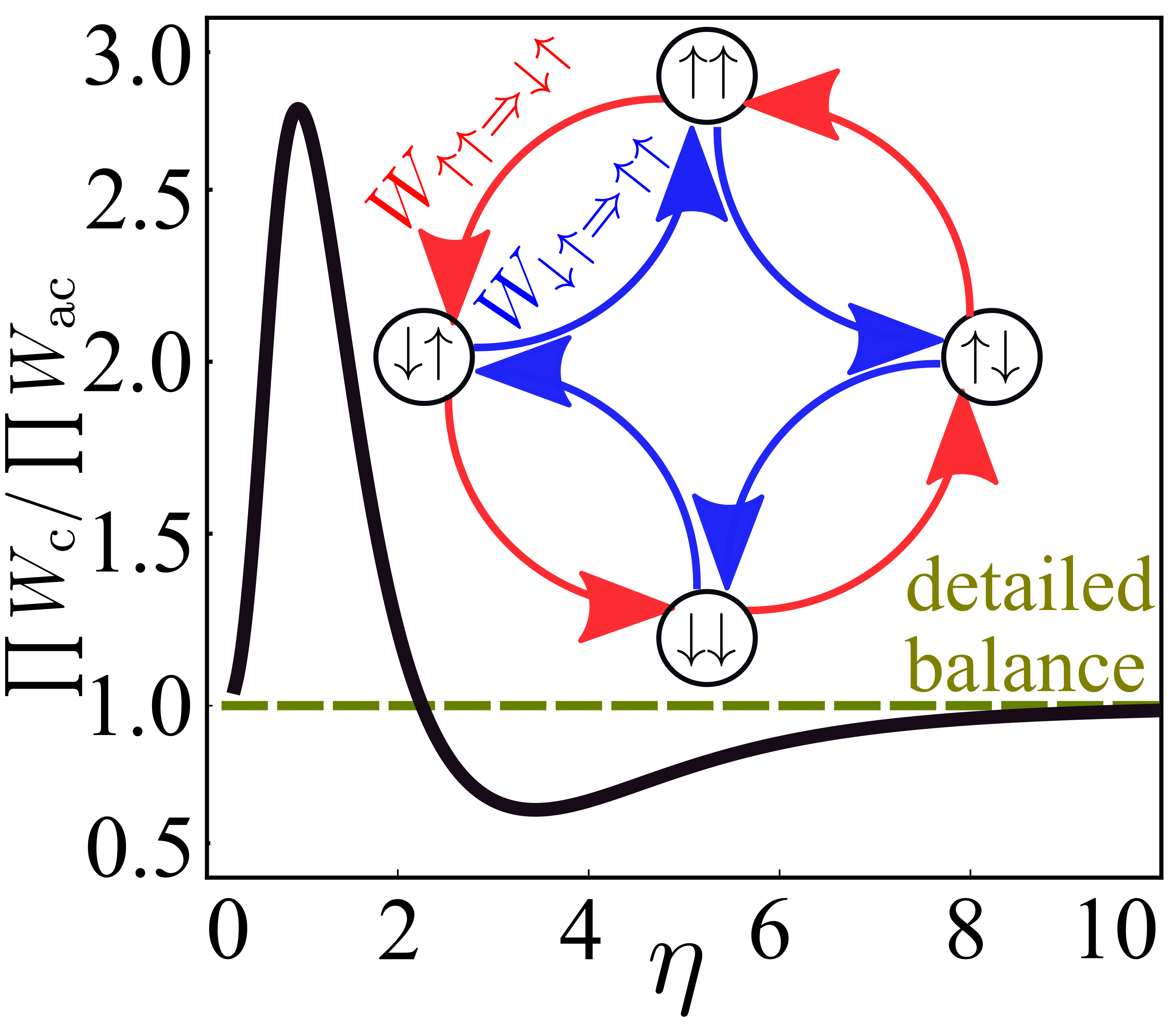}
\caption{\textbf{Non-equilibrium regime.} Kolmogorov criterion in a two-spin subspace; the  displayed quantity is the ratio of the product $\prod W_\mathrm{c}$ of the four rates encountered when performing the loop clockwise (blue arrows) divided by the analogous counter-clockwise (red arrows) product $\prod W_\mathrm{ac}$, plotted as a function of $\eta$. This ratio is $1$ (signalling detailed balance conditions) only for large and very small values of $\eta$. The intermediate point $\eta \approx 2$ where $\prod W_\mathrm{c}=\prod W_\mathrm{ac}$ can be safely ignored for the following reason: for systems with more than two spins there are multiple loops in configuration space and Kolmogorov's criterion is never satisfied simultaneously for all of them (except in the extremal limits $\eta \to 0$ and $\eta \to \infty$) and the dynamics does not obey detailed balance. The parameters are:  $\omega=1$, $\Omega=0.1$, $\kappa=1$ and $g_0=4$. 
}
\label{f:fig3}
\end{figure}
To obtain approximate expressions for the transition rate we treat it as a function of the energy difference $\Delta E$, which we now regard as a continuous parameter. Furthermore, we note that, for sufficiently small $\eta$, the integrand defining $W_{\vs \rightarrow \vs'}$ is rapidly suppressed for $\tau >0$ due to a fast initial increase of $f(\tau) \approx 2(1-\cos \tau)/\eta$. Thus, the integral is dominated by the contribution close to $\tau = 0$. Hence, one can expand all arguments in powers of $\tau$ (see Appendix). Setting for simplicity $\omega = \kappa = 1$ and keeping for brevity only the leading orders in $\eta \to 0$, one obtains $\tau + f(\tau) \approx \tau^2/\eta - \tau^3/6$, $\nu \approx 2$ and $s(\tau) \approx - \tau + \tau^3/6$. This implies that the suppression of the integrand occurs on a timescale $\tau \sim \sqrt{\eta/(4g_i^2)}$, whereas the cosine term oscillates with a frequency which is approximately $\Gamma_i = 4(\Delta E + g_i^2)$. We thereby identify (i) a regime of ``small energy jumps'', where $\Gamma_i^2 \ll 4g_i^2/\eta$ and (ii) a ``large energy jumps'' one with $\Gamma_i^2 \gg 4g_i^2/\eta$. In case (i), we obtain (see Appendix)
\begin{eqnarray}
W_i (\Delta E) 
&\approx& \frac{\Omega^2}{2} \sqrt{\frac{\pi\eta}{g_i^2}} e^{-\frac{\eta}{g_i^2} (\Delta E + g_i^2)^2},
\label{eq:approx}
\end{eqnarray}
where the index $i$ reminds us of which spin is being flipped \footnote{Note, that in this notation if $W_{\vs \rightarrow \vs'} \equiv W_i(\Delta E)$ then $W_{\vs' \rightarrow \vs} \equiv W_i(-\Delta E)$).}. The rate reaches its maximum when $\Delta E = - g_i^2 \leq 0$ and, in general, $W_i(\abs{\Delta E}) < W_i (-\abs{\Delta E})$. This means that spin flips which lower the energy are favored, suggesting that the dynamics enacts a form of cooling. Case (ii) can be analyzed using the asymptotic expansion of Fourier integrals \cite{Dai1992} (see also Appendix). To leading order this yields a power-law decay $W_i (\Delta E) \approx 8 \Omega^2 g_i^2 \Gamma_i^{-4}$, which shares the same ``cooling'' properties. A numerical integration suggests that $W_i(\abs{\Delta E}) < W_i (-\abs{\Delta E})$ holds also in between the asymptotic cases (i) and (ii).

To shed further light on the cooling dynamics we analyze this asymmetry of the rates through the ratio $R_i(\Delta E) = W_i(\Delta E) / W_i(-\Delta E)$. This is depicted in Fig.~\ref{f:fig2}(c,d) as a function of $\Delta E$ and $\eta$, respectively. In regime (i) we have $R_i(\Delta E) \approx e^{-4\eta \Delta E}$, which implies a thermal dynamics with an effective inverse temperature $\beta_{\rm eff} = 4 \eta$. Note that the r.h.s.~has no dependence on the index $i$, implying the existence of a unique, well-defined temperature for all spin-flip processes. In Fig.~\ref{f:fig2}(e) we display the ratio $\log[R_i(\Delta E)] / \Delta E$ for different values of $\Delta E$ and show that different curves collapse to a single (negative) inverse temperature $-\beta_\mathrm{eff}$ up to the edge of case (i). At $\eta=0$ we have $\beta_{\rm eff}$ approaches zero, leading to an infinite-temperature dynamics. This is reasonable, since in this limit the bosonic gain rate approaches the loss rate. This implies the population of arbitrarily high Fock states, effectively heating the bosons. The latter then act as a high-temperature bath on the spins. If, on the other hand, $1/\beta_{\rm eff}$ remains small or comparable with the energy gap from the ground states of (\ref{cost}) --- which on average is of order $g^2$, meaning $4 \eta g^2 \geq 1$) --- an effective low-temperature dynamics is realized.

In case (ii), the ratio $R_i(\Delta E) \approx (\Delta E - g_i^2)^4/(\Delta E + g_i^2)^4$ tends to increase towards $1$ as $\Delta E$ grows. Typically, the available $\Delta E_i$s populate both range (i) and (ii), implying the presence of type (ii) processes which do not follow the same low-temperature rules obeyed by the ``small-jump'' ones. Provided the number of spins $N$ is not too large, these non-thermal processes constitute, however, a small perturbation for the following two reasons: first, the distribution of energy jumps is peaked around $0$, implying that, if the parameters are adequately chosen, most jumps lie in regime (i). Second, since the rates are decreasing functions of $\abs{\Delta E + g_i^2}$, type (ii) processes occur at smaller rates than the type (i), thermal ones. A numerically exact analysis of the classical master equation for $N=10$, displayed in Fig.~\ref{f:fig2}(b), indeed shows that the effect of the non-thermal processes is sufficiently weak to avoid having a significant population of high-energy states in the long-time limit. The statistical energy distribution tends instead to become concentrated on low-energy configurations, highlighting a clear bias of the dynamics towards cooling, as compared e.g., to the $\eta \gg 1$ case in panel (a).

\emph{Breakdown of detailed balance ---} Outside the thermal regimes the dynamics is not an equilibrium one, i.e.~it does not obey detailed balance. This can be proved via Kolmogorov's criterion \cite{Zia_2007} which we analyze for the loop formed in the configuration space of a two-spin system (see Fig.~\ref{f:fig3}): $(\uparrow \uparrow) \rightarrow (\uparrow \downarrow) \rightarrow (\downarrow \downarrow) \rightarrow (\downarrow \uparrow) \rightarrow (\uparrow \uparrow)$. To this end we investigate the ratio between the product of the rates for the clockwise (blue arrows) cycle and the corresponding product for the counter-clockwise (red arrows) one. This ratio goes to $1$ when $\eta \to \infty$ and  also when $\eta \to 0$, signalling the emergence of the infinite-temperature dynamics. For different values $\eta$ the ratio is typically different from one, which indicates the persistence of probability currents in the stationary state and the absence of detailed balance.

\emph{Conclusions ---} We have studied a disordered dissipative spin-boson system in the limit of strong coupling and weak driving, which can for example be implemented on trapped ion quantum simulators. Many aspects of the emerging physics can be understood in terms of a disordered fully-connected Ising model whose state evolves according to a rate equation. In general the dynamics violates detailed balance, and the system is thus out of equilibrium. However, we could identify parameter regimes in which the evolution is effectively thermal. Among them is one where predominantly low-energy configurations are populated,which mimics the action of a low-temperature dynamics. In the future it would be interesting to see whether this effective cooling mechanism permits to access low-energy states or even ground states of complex spin networks. This might open an elegant way for encoding and solving computationally hard problems \cite{Barahona_1982} or associative memories \cite{Hopfield:1982} through Ising energy functions and an appropriate (thermal) dynamics on quantum simulators.

\emph{Acknowledgments---} The research leading to these results has received funding from the European Research Council  under  the  European  Unions  Seventh  Framework  Programme (FP/2007-2013)/ERC Grant  Agreement No. 335266  (ESCQUMA), the EPSRC Grant No.EP/R04340X/1 via the QuantERA project “ERyQSenS and from the University of Nottingham through a Nottingham Research Fellowship (M.M.). P.R. acknowledges funding by the European Union through the H2020 - MCIF No. $766442$. I.L. gratefully acknowledges funding  through the Royal Society Wolfson Research Merit Award.

\bibliographystyle{apsrev4-1}

\bibliography{DM_bib}

\pagebreak

\widetext
\begin{center}
        \textbf{\large Appendices for "Dynamics of strongly coupled disordered dissipative spin-boson systems"}
\end{center}

\setcounter{equation}{0}
\setcounter{figure}{0}
\setcounter{table}{0}
\makeatletter
\renewcommand{\theequation}{S\arabic{equation}}
\renewcommand{\thefigure}{S\arabic{figure}}

\renewcommand{\bibnumfmt}[1]{[S#1]}


\section{Derivation of the rates}

In this section we provide details on the derivation of Eq.(3) in the main text. We firstly consider the evolution of the state $\rho(t)$ as $\dot{\rho}=\lind_{0}(\rho)+\lind_{1}(\rho)$, with $\lind_{0}=\lind-\lind_{1}$ and $\lind_{1} (\cdot)=-i\Omega[\sum_{i}\hat{\sigma}^{x}_{i}, \cdot]$. Secondly, we assume the stationary state of $\lind_{0}$ of the form $\rho_{\mathrm{stat}}=\sum_{\vs}p_{\vs}\rho_{\vs}\ket{\vs}\bra{\vs}$, where $\ket{\vs}=\lbrace \sigma_{1},...,\sigma_{N} \rbrace$, with $\sigma_{i}=\pm 1$ and $\hat{\sigma}_{i}^{z}\ket{\vs}=\sigma_{i}\ket{\vs}$, $p_{\vs}$ are a set of classical probabilities and $\rho_{\sigma}$ is the corresponding bosonic state, that we assume to be a gaussian state. Considering $\lind_{1}$ perturbatively with respect to $\lind_{0}$, and projecting the dynamics onto the stationary manifold of $\lind_{0}$, we exploit the Nakajima-Zwanzig formalism to write the evolution of the spin as 
\begin{equation}\label{s.NZ}
\begin{split}
 P\dot{\rho}^{\mathrm{spin}}(t) =&  \tr_{B} \int_{0}^{+\infty} dt' P \lind_{1}e^{\lind_{0}T'}\lind_{1}\rho_{\mathrm{stat}}(t)=\\
=& \Omega^{2}\sum_{\lbrace \vs \rbrace}p_{\vs}(t)\sum_{i}\int_{0}^{+\infty} dt' \sum_{j=\pm} \tr_{B}\left[ e^{\V_{\vs,i}^{j}t'}(\rho_{\vs}) \right] \times \\
& \times \left( \hat{\sigma}_{i}^{x}\ket{\vs}\bra{\vs}\hat{\sigma}^{x}_{i} - \ket{\vs}\bra{\vs} \right),
\end{split}
\end{equation}
where $P\dot{\rho}^{\mathrm{spin}}(t)=\tr_{B}[\dot{\rho}_{\mathrm{stat}}(t)]=\sum_{\vs}\dot{p}_{\vs}(t)\ket{\vs}\bra{\vs}$, $\tr_{B}$ is the partial trace over the boson, and we have defined the spin-configuration dependent superoperators 
$$\V_{\vs,i}^{\pm}(\cdot)= -i\omega [\hat{a}^{\dagger}\hat{a},\cdot ] + \dg \left(\cdot \right)+ \dk \left(\cdot \right)- ig_{i}\mathcal{M}_{i}\left[(\hat{a}^{\dagger}+\hat{a}), \cdot \right]\pm i g_{i} \sigma_{i} \left\lbrace (\hat{a}^{\dagger}+\hat{a}),\cdot \right\rbrace\,,$$ with $\mathcal{M}_{i}=\sum_{l \neq i}g_{l}\sigma_{l}$, and $\dg$, $\dk$ the dissipative terms representing cooling and heating, respectively. By projecting Eq.(\ref{s.NZ}) on a state $\ket{\vs'}$, the dynamics reduces to the evolution of the classical probabilities  ruled by a master equation whose general form is the following
\begin{equation}\label{e.spindynamics}
\dot{p}_{\vs}=\sum_{\vs'} \left( W_{\vs' \rightarrow \vs}p_{\vs'}- W_{\vs \rightarrow \vs'}p_{\vs} \right)\,,
\end{equation}
where $W_{\vs \rightarrow \vs'}=\Omega^2 \int_{0}^{+\infty} dt \sum_{j=\pm} \tr_{B}\left[ e^{\V_{\vs,i}^{j}\tau}(\rho_{\vs}) \right]$, is the transition rate for the switching  $\vs \rightarrow \vs'$ and it allows only single spin-flip processes.   

We can now go ahead in evaluating explicitly the expression for the rates. Exploiting the superoperator's properties, we can write $e^{\V_{\vs,i}^{\pm}}(\rho_{\vs})=(e^{\V_{\vs,i}^{\pm, *}}\mathbb{1})\rho_{\vs}$, with $\V_{\vs,i}^{\pm, *}$ the adjoint superoperator of $\V_{\vs,i}^{\pm}$ and $\mathbb{I}$ the identity operator. It is worth noticing that the identity operator can be expressed in terms of a generalised displacement operator of field coherent states as $\mathbb{I}=\hat{D}(0)$, where $\hat{D}(\tau)=e^{\alpha(\tau)\hat{a}^{\dagger}-\beta(\tau)^{*}\hat{a}}e^{-\gamma(\tau)}$ with $\alpha(0)=\beta(0)=\gamma(0)=0$. We then verify that the displacement operator $\hat{D}(0)$ is mapped into the generalised one $\hat{D}(\tau)$ by applying  the adjoint superoperator $\V_{\vs,i}^{\pm}$: indeed, by considering the differential equation $\frac{d}{d\tau}\left[ \hat{D}_{\vs,i}^{\pm}(\tau)\right]=\V_{\vs,i}^{\pm *}\left[ \hat{D}_{\vs,i}^{\pm}(\tau) \right]$ we obtain the solutions for the functions $\alpha^{+}_{\sigma_{i}}(\tau)=\alpha^{-}_{-\sigma_{i}}(\tau)$, $\beta^{+}_{\sigma_{i}}(\tau)=\beta^{-}_{-\sigma_{i}}(\tau)$, $\gamma^{+}_{\sigma_{i}}(\tau)=\gamma^{-}_{-\sigma_{i}}(\tau)$.  For initial conditions $\alpha_{\sigma_{i}}^{\pm}(0)=\beta_{\sigma_{i}}^{\pm}(0)=\gamma_{\sigma_{i}}^{\pm}(0)=0$, we get
\begin{eqnarray}
\begin{split}
& \alpha^{+}_{\sigma_{i}}(\tau)= [\beta_{\sigma_{i}}^{+}(t)]^{*}=\frac{i4g_{i}\sigma_{i}}{\omega(\eta-2i)}\left[1- e^{(i  - \frac{\eta}{2})\tau} \right],\\
& \gamma_{\sigma_{i}}^{+}(\tau)=  \frac{2g_{i}^{2}\nu}{\omega^{2}\eta}\left[f_{1}(\tau) +\tau \right] + \frac{ig_{i}\sigma_{i}\mathcal{M}_{i}}{\omega^{2}(\eta^{2}+4)}[s(\tau)+\tau]\, ,\\
& f_{1}(\tau)=\frac{1-e^{\eta \tau}}{\eta}-\frac{4\eta[1-e^{-\frac{\eta}{2}\tau}\cos{(\tau)}]-8e^{-\frac{\eta}{2}\tau}\sin(\tau)}{\eta^{2}+4},\\
& s(\tau) = \frac{  4\eta\left[ e^{-\frac{\eta}{2}\tau}\cos(\tau)-1 \right] + \left(\eta^{2}-4 \right)e^{-\frac{\eta}{2}\tau}\sin(\tau)  }{ \eta^{2}+4}\,,
\end{split}
\end{eqnarray}
where we have defined the dimensionless time $\tau=t \omega$, and $\eta=(\gamma-\kappa)/\omega$,  $\theta = \kappa/\gamma \in [0,1)$, and 
\begin{equation}
	\nu= \frac{4(1+\theta)\eta}{(\eta^2+4)(1-\theta)} = \frac{4\left(  2\frac{\kappa}{\omega}+\eta \right)}{\eta^2+4}.
\end{equation}

The previous steps allow us to write the partial trace over the boson as $\tr_{B}[e^{\V_{\vs,i}^{\pm}}(\rho_{\vs})]=e^{-\gamma_{\sigma_{i}}^{\pm}(\tau)}\tr_{B}[\hat{D}^{\pm}_{\vs,i}(\tau)\rho_{\vs}] $. 
We recall that the bosonic state $\rho_{\vs}$ has been assumed to be a gaussian state. In this case, we recognise the quantity $\tr_{B}[\hat{D}^{\pm}_{\vs,i}(\tau) \rho_{\vs}] $  as the \textit{characteristic function} $\chi^{\pm}_{\vs,i}(\tau)$ of the state $\rho_{\vs}$. The expression of the characteristic function for a generic gaussian state $\rho_{G}$ reads  
\begin{equation}\label{s.chi}
\chi_{\rho_{G}}[\alpha(\tau)]= e^{-\frac{1}{4}\vec{\alpha}^{T}(\tau)\Sigma\vec{\alpha}(\tau)+\alpha(\tau)\braket{\hat{a}^{\dagger}}_{G}-\alpha^{*}(\tau)\braket{\hat{a}}_{G}},
\end{equation}
where $\vec{\alpha}^{T}=(\alpha(\tau), \alpha^{*}(\tau))$, $\braket{\cdot}_{G}$ is the expectation value performed over the state $\rho_{G}$, and $\Sigma$ represent the covariance matrix which reads
\begin{equation}\label{s.covariance}
2\begin{pmatrix}
-(\braket{\hat{a}^{2}}_{G}-\braket{\hat{a}}^{2}_{G}) && \frac{1}{2}(\braket{\hat{a}^{\dagger}\hat{a}}_{G}+\braket{\hat{a}\hat{a}^{\dagger}}_{G})-\braket{\hat{a}^{\dagger}}_{G}\braket{\hat{a}}_{G}\\
\frac{1}{2}(\braket{\hat{a}^{\dagger}\hat{a}}_{c}+\braket{\hat{a}\hat{a}^{\dagger}}_{G})-\braket{\hat{a}^{\dagger}}_{G}\braket{\hat{a}}_{G} && -(\braket{\hat{a}^{\dagger \; 2}}_{G}-\braket{\hat{a}^{\dagger}}^{2}_{G})
\end{pmatrix}.
\end{equation}
By applying the definition \eqref{s.chi}, with expectation values of the operators obtained considering the lindblad operator $\lind_{0}$, we get
\begin{equation}
\chi^{\pm}_{\vs,i}(\tau)= \exp{\left\lbrace-\frac{1+\theta}{2(1-\theta)}|\alpha_{\sigma_{i}}^{\pm}(\tau)|^{ 2}+ \frac{2ig_{i}\mathcal{M}}{\omega}\left(\frac{\alpha_{\sigma_{i}}^{\pm}(\tau)}{\eta-2i}+\frac{\alpha_{\sigma_{i}}^{\pm *}(\tau)}{\eta+2i} \right)\right\rbrace},
\end{equation}
where $\mathcal{M}=\sum_{l}g_{l}\sigma_{l}$. Thus, the expression of the rate reads
\begin{equation}
\begin{split}
W_{\vs \rightarrow \vs'}= & \frac{\Omega^2}{\omega} \int_{0}^{+\infty} d\tau \sum_{j=\pm}e^{-\gamma^{j}_{\sigma_{i}}(\tau)} \chi_{\vs,i}^{j}(\tau)= \frac{2\Omega^{2}}{\omega}\int_{0}^{\infty}\! d\tau e^{-\frac{2g_{i}^{2}\nu}{\omega^{2}}\left[ f(\tau)+\tau \right] } \begin{medsize}  \cos{\left[ 16 \frac{\Delta E_{i} \tau - g_{i}^{2}s(\tau) }{\omega^{2}(\eta^{2}+4)} \right]} \end{medsize}\,, \\
&f(\tau) = \frac{-2\eta^2+8}{\eta \left(\eta^2+4\right)} \left[ 1-e^{-\frac{\eta}{2}\tau}\cos( \tau) \right]-\frac{8 e^{-\frac{\eta}{2}\tau} }{\eta^{2}+4}\sin( \tau)\,, \\
\end{split}
\end{equation}
where  $\Delta E_i~= g_i \sigma_i\sum_{l\neq i}g_{l}\sigma_{l}$ retains the dependence on the spin configuration.

\section{Approximate expression of the rate for $\eta$ small}

For sufficiently small $\eta \lesssim 1$, the exponent appearing in $W_{\vs \rightarrow \vs'}$ is dominated by
\begin{equation}
	f(\tau) \approx \frac{2}{\eta} (1 - \cos( \tau)),
\end{equation}
implying that the integrand is quickly suppressed as $\tau$ grows. We therefore perform an expansion around $\tau = 0$ which yields
\begin{equation}
	\tau + f(\tau) = (\eta^2+4) \left(  \frac{1}{4\eta} \tau^2 - \frac{1}{24} \tau^3 - \frac{4-\eta^2}{192 \eta} \tau^4 \right) + O(\tau^5)
	\label{eq:fexp}
\end{equation}
and
\begin{equation}
	s(\tau) = - \tau + \frac{\eta^2 + 4}{24} \tau^3 - \eta \frac{\eta^2 + 4}{96} \tau^4 + O(\tau^5).
\end{equation}
From the first term in Eq.~\eqref{eq:fexp} we see that the integrand is strongly suppressed on scales $\tau \gtrsim \sqrt{\eta}$. Noticing that in the Taylor expansion of $\tau + f(\tau)$ odd coefficients are finite, whereas even ones are $O(1/\eta)$ for $\eta \to 0$ and introducing the rescaled integration variable $z = \tau \sqrt{\eta}$ we see that higher orders are perturbations of order $O(\eta^{n+1/2} z^{2n+1}, \eta^{n-1} z^{2n})$ and can be neglected. Similar considerations can be applied to $s(\tau)$, which can be therefore also approximated with its leading order $-\tau$. In the following, for simplicity we set $\omega = 1$, remembering that our ``energy'' $\Delta E_i$ is actually measured by construction in units of $\omega^2$. Additionally, we introduce the shorthand
\begin{equation}
	\Gamma_i = \frac{16}{\eta^2 + 4} (\Delta E_i + g_i^2),
\end{equation}
so that, by keeping only the lowest orders of the expansion in $\tau$, we can approximate our rate as
\begin{equation}
	W_{\vs \rightarrow \vs'} \approx 2\Omega^{2} \int_{0}^{\infty}\! d\tau \, e^{-2\frac{g_{i}^{2}\left( 2\kappa + \eta \right)}{\eta}\tau^2  }  \cos{\left( \Gamma_i \tau  \right)} ,
	\label{eq:intermediate}
\end{equation}
which can be integrated to give the closed expression
\begin{equation}
	W_{\vs \rightarrow \vs'} \approx  \Omega^2 \sqrt{\frac{\pi \eta }{2g_i^2 \left( 2\kappa + \eta \right)}} e^{- \frac{\eta \Gamma_i^2}{8 g_i^2 \left( 2\kappa + \eta \right)}}.
	\label{eq:Gauss1}
\end{equation}
It is worth remarking that the exponent can be rewritten as
\begin{align}
	- \frac{\eta \Gamma_i^2}{8 g_i^2 \left( 2\kappa + \eta \right)} = \frac{128 \eta}{(2\kappa + \eta)(\eta^2 + 4)^2} \left[- \frac{(\Delta E_i + g_i^2)^2}{4g_i^2}  \right] = \nonumber \\
	 = \frac{128 \eta}{(2\kappa + \eta)(\eta^2 + 4)^2} \left[ E(\vec{\sigma}) -  \frac{\sum_j g_j^2}{4}   \right] =  \beta_{\rm eff} E(\vec{\sigma}) - const.\, ,
\end{align}
highlighting the ``thermal'' structure of the rates.
Note that, in order to obtain the approximation \eqref{eq:intermediate}, we have assumed that we can resum the Taylor expansion of the original cosine to the function $\cos(\Gamma_i \tau)$, whose series only coincides with the former up to $O(\tau^2)$. This is only valid as long as the cosine does not oscillate significantly before the other Gaussian term suppresses the integrand; in other words, Eq.~\eqref{eq:Gauss1} should be valid up to values of $\Gamma_i$ of the order of $\sim 1/\sqrt{\eta}$. Since we wish to understand the behavior of the rates as functions of the energy difference $\Delta E_i$ without restrictions imposed by the other parameters (like $\eta$), we need to account for energies which exceed this range. To do this, we extract the asymptotic behavior of the rate for $\Gamma_i \to \infty$. We start by rewriting the integrand in $W$ as
\begin{align}
	I(\Gamma_i, \tau)&  \equiv {\mathrm{Re}} \left\{  e^{-2\frac{g_{i}^{2}\nu}{\omega^{2}}\left( f(\tau)+\tau \right) }   e^{i \Gamma_i \tau  - 16 i \frac{g_i^2 }{\eta^2 + 4} (s(\tau) + \tau)}        \right\} = \\
	& = {\mathrm{Re}} \left\{  A(\tau) e^{i\Gamma_i \tau} \right\}.
\end{align}
We now use the result that, if the function $A$ admits a small $\tau$ expansion
\begin{equation}
	A(\tau) = \sum_{n=0}^\infty a_n \tau^n, 
\end{equation}
then asymptotically in the limit $\Gamma_i \to \infty$ one finds
\begin{equation}
	\int_{0}^{\infty}\! d\tau \, A(\tau) e^{i\Gamma_i \tau} = \sum_{n=0}^\infty i^n \, n! \, a_n   \Gamma_i^{-n-1}.
\end{equation}
The leading term in this expansion corresponds to the lowest $n$ for which one finds a non-vanishing real part. In particular, we note that ${\mathrm{Re}} [a_n i^{n+1}]$ equals $(-1)^{l+1} {\mathrm{Re}} [a_{2l+1}]$ if $n = 2l+1$ is odd, and $(-1)^{l+1}{\mathrm{Im}} [a_{2l}] $ if $n = 2l$ is even. For our function we find
\begin{subequations}
\begin{align}
	a_0 &= 1, \\
	a_1 &= 0, \\
	a_2 &= -2 g_i^2 \frac{2\kappa + \eta}{\eta}, \\
	a_3 & = \frac{g_i^2}{3} \left[ 2 \kappa + \eta - 2 i \right].
\end{align}
\end{subequations}
The leading behavior in the large $\Gamma_i$ limit is therefore determined by ${\mathrm{Re}}[a_3]$, implying
\begin{equation}
	W_{\vs \rightarrow \vs'} = 2\Omega^2 \int_0^\infty d\tau \, I(\Gamma_i, \tau) \approx 4 \Omega^2 g_i^2 (2\kappa + \eta) \left[ \frac{\eta^2 + 4}{16(\Delta E_i + g_i^2)}  \right]^4 .
	\label{eq:largeE}
\end{equation}

To provide a very crude estimate of where the change from the two regimes characterized by Eqs.~\eqref{eq:Gauss1} (``small $\Gamma_i$'') and \eqref{eq:largeE} (``large $\Gamma_i$'') occurs, we evaluate the point where the two asymptotic expressions cross (for $\eta$ sufficiently small): setting
\begin{equation}
	4 g_i^2 (2\kappa + \eta) \Gamma_i^{-4} = \sqrt{\frac{\pi \eta }{2g_i^2 \left( 2\kappa + \eta \right)}} e^{- \frac{\eta \Gamma_i^2}{8 g_i^2 \left( 2\kappa + \eta \right)}}
\end{equation}
we find
\begin{equation}
	\Gamma_i^2 \approx \frac{4g_i^2 (2\kappa + \eta)}{\eta}  \left[ \log A + 4\log\left(  \frac{1}{2} \log A \right)   \right] ,
\end{equation}
where $A = 32 g_i^2 \pi (2\kappa +\eta) \eta^{-3}$.

\end{document}